# On an Efficient Marie Curie Initial Training Network

**Ali Dinler[(1)], Cengis Hasan[(2)], Kamil Orucoglu[(3)] and Robert W. Barber[(4)]**

ali.dinler@stfc.ac.uk, cengis.hasan@inrialpes.fr, koruc@itu.edu.tr and robert.barber@stfc.ac.uk

[(1), (4)] Centre for Microfluidics and Microsystems Modelling, STFC Daresbury Laboratory, Warrington WA4 4AD, United Kingdom

[(2)] INRIA, CITI Laboratory, Av. des Arts 69621, Villeurbanne Cedex, Lyon, France

[(3)] Istanbul Technical University, Department of Mathematics, 34469, Maslak Istanbul, Turkey

**ABSTRACT**

Collaboration in science is one of the key components of world-class research. The European Commission supports collaboration between institutions and funds young researchers appointed by these partner institutions. In these networks, the mobility of the researchers is enforced in order to enhance the collaboration. In this study, based on a real Marie Curie Initial Training Network, an algorithm to construct a collaboration network is investigated. The algorithm suggests that a strongly efficient expansion leads to a star-like network. The results might help the design of efficient collaboration networks for future Initial Training Network proposals.

**INTRODUCTION**

Social network models are important for investigating collaboration and communication between people or organizations. Newman [1] has shown that the connectedness and closeness are useful definitions when analysing scientific collaboration. In addition, the average shortest path, the diameter (*i.e.* the maximum shortest path) and the density of such networks are useful measures for the comparison of networks or comparison of the partners within a network. However, cost and payoff (benefit) definitions are necessary for the investigation of the overall efficiency of the network. Jackson and Wolinsky [2], in the context of cooperative game theory, have introduced the connections model and proven that a unique *star network*[1] maximizes the overall payoff of the network, in the case where the cost of maintaining a link is higher than the value of a link but low enough that overall payoff keeps increasing with new connections.

In this paper, an algorithm for designing an efficient Marie Curie Initial Training has been investigated. The collaborative network is assumed to consist of fourteen partner institutions and seventeen early stage researchers (ESRs). Each partner hires at least one ESR and each ESR must visit two other partners. The lengths of the visits vary depending on the career development plan of each ESR. The strength of collaboration between any two partners is determined by the length of these visits; a longer visit implies more collaboration. Therefore the distance between any two partners is defined as the inverse of the total length of visits of the ESR(s) travelling from one partner to another. The problem considered in this study differs from the connections model of Jackson and Wolinsky [2] in that: (a) there is a primary network consisting of four founding partners, (b) the distances between partners vary according to the length of the ESR visits, and (c) there are three different payoff values that can be received from a created link.

## DEFINITIONS AND PROBLEM DESCRIPTION

We have fourteen partners $\{P_1,...,P_{14}\}$. Each edge, $ij$, between partners $P_i$ and $P_j$ represents an undirected link and the *distance* between any two partners is defined as the inverse of the total length of the ESR visits. For example, in the case when the ESR of $P_1$ visits $P_4$ for three months and the ESR of $P_4$ visits $P_1$ for fifteen months, the distance between $P_1$ and $P_4$ becomes 1/(15+3)=0.05556.

The network consists of eight experimental and six computational partners. The partners $P_1, P_2, P_3, P_7, P_8, P_9, P_{11}$ and $P_{14}$ are conducting experimental research; and partners $P_4, P_5, P_6, P_{10}, P_{12}$ and $P_{13}$ are conducting computational research. The cost of creating a link is assumed to be equal throughout the network. However, the payoff of a created link depends on whether the link is between two computational (c–c) or two experimental (e–e) partners or between an experimental and a computational partner (e–c). The payoff, $\delta$, is assumed to be highest for e–c, and lowest for c–c, i.e. $\delta^{e-c} > \delta^{e-e} > \delta^{c-c}$.

The aim of this study is to define an appropriate efficiency concept, propose an algorithm to find an efficient network, and then examine the final structure of the network from the point of view of both the partners and ESRs.

---

[1] A star network is a star-shaped network with a central player such that all other players are connected to this central player.

We start with the network of four founding partners, denoted by $g_f$. Each founding partner hires one ESR. The length of visits of ESRs within the $g_f$ have been decided at the beginning of the project, as shown in Table 1. Thus the distances between founding partners are already known, as shown in Figure 1. The distances are calculated as the inverse of the total visits of ESRs between any two partners. As new partners are linked sequentially, the network, $g$, expands in a way that an efficient network is constructed.

|  | $P_1$ | $P_2$ | $P_3$ | $P_4$ |
|---|---|---|---|---|
| $ESR_1 (P_1)$ | 0 | 0 | 8 | 3 |
| $ESR_2 (P_2)$ | 2 | 0 | 16 | 0 |
| $ESR_3 (P_3)$ | 8 | 8 | 0 | 0 |
| $ESR_4 (P_4)$ | 15 | 0 | 3 | 0 |

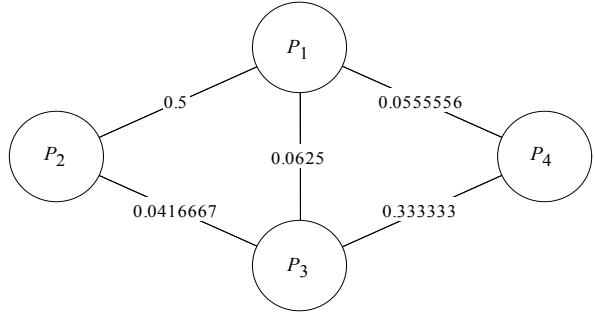

Table 1. *Length (months) of visits of the early stage researchers (ESRs) within the founding network, $g_f$. $ESR_1$, $ESR_2$, $ESR_3$ and $ESR_4$ have been appointed by partners, $P_1$, $P_2$, $P_3$ and $P_4$, respectively.*

Figure 1. *Graph of the network of founding partners. The numbers along the edges represent the distances. The distance is defined as the inverse of the length of total visits of ESRs between any two partners.*

The net *individual payoff* that $P_i$ receives from the network $g$ is[2]

$$u_i = \sum_{\substack{j \neq i \\ j=1}}^{N} \frac{\delta_{ij}}{d_{ij}} - \sum_{j=1}^{M} c_{ij} \qquad (1)$$

where $d_{ij}$ is the shortest distance between $P_i$ and $P_j$, $N$ is the number of partners, $c_{ij}$ is the cost of creating a link and $M$ is the number of direct links to $P_i$. The payoffs are specified as $\delta^{e-c} = 3$, $\delta^{e-e} = 2$ and $\delta^{c-c} = 1$. The cost of a link is assumed to be equal throughout the network, (i.e. $c_{ij} = 1$). According to the definition (1), each partner $P_i$ benefits from indirect links while the cost only comes from the creation of direct links between $P_i$ and its adjacent partners.

---

[2] The definition in Eq. (1) is inspired from Jackson and Wolinsky [2]. However, in this study, it is based on the shortest distances in a different manner and there are three different payoff ($\delta$) values.

The *value of the network g* is [2]

$$v(g) = \sum_{i=1}^{N} u_i \qquad (2)$$

A network is referred to as *strongly efficient* [2] if its value is maximized on the set of all possible networks.

**Definition:** Let $g_f$ be the network of founding partners. The *strongly efficient expansion* is the expansion of $g_f$ by linking *m* new partners so that the value of the network is maximized on the set of all possible expansions.

The problem discussed here is to find a strongly efficient expansion of the founding network $g_f$ according to the length of visits shown in Tables 1 and 2. There are three new partners ($P_5$, $P_6$ and $P_7$) that hire two ESRs while the other partners hire only one ESR.

| $P_5$ | $ESR_5$ | 9, 4 |
|---|---|---|
|  | $ESR_6$ | 8, 6 |
| $P_6$ | $ESR_7$ | 8, 4 |
|  | $ESR_8$ | 10, 4 |
| $P_7$ | $ESR_9$ | 9, 4 |
|  | $ESR_{10}$ | 6, 5 |
| $P_8$ | $ESR_{11}$ | 8, 7 |

| $P_9$ | $ESR_{12}$ | 10, 4 |
|---|---|---|
| $P_{10}$ | $ESR_{13}$ | 9, 5 |
| $P_{11}$ | $ESR_{14}$ | 10, 3 |
| $P_{12}$ | $ESR_{15}$ | 8, 5 |
| $P_{13}$ | $ESR_{16}$ | 8, 4 |
| $P_{14}$ | $ESR_{17}$ | 6, 3 |

Table 2. *Lengths (months) of visits of ESRs appointed by new partners. Each ESR must visit two other partners for the periods of time shown in the table. For example, $ESR_5$ is going to visit a partner for 9 months and visit another partner for 4 months.*

## ALGORITHM

The algorithm for finding a strongly efficient expansion requires an exhaustive search on the set of all possible networks (without any prior knowledge leading to the solution). Even for ten new partners, this search is prohibitive. Carayol *et al.* [3], for a similar problem, suggested a genetic algorithm to construct an efficient communication network. However, in this paper, we use a simple heuristic algorithm.

Initially, all new partners are arranged in descending order according to the total mobility of their ESRs. In other words, the partner which sends its ESR(s) for longer periods of time to the other partners links first. The algorithm starts with $g_f$ and links each new partner one by one to the most appropriate partners. It is assumed that the partners hiring two ESRs and sending them for longer times are more eager to collaborate and are more likely to host other ESRs. As they link first, other partners are able to link to these enthusiastic partners.

**Algorithm:**

1. Calculate all shortest distances within the network of founding partners, $g_f$. Find the shortest distance matrix, $M_{g_f}$ (which is a $4 \times 4$ matrix).
2. Divide each shortest distance by the appropriate payoff $(\delta)$ value and find the weighted shortest distance matrix, $W_{g_f}(i,j) = M_{g_f}(i,j)/\delta_{ij}$.
3. Do i=5:17
    4. Choose any two partners in the network:
    
    Link the partner of $ESR_i$ to these two partners. Calculate all shortest distances and divide each shortest distance by the appropriate payoff $(\delta)$. Find the average of all the elements of the weighted shortest distance matrix. Keep the expanded network with the minimum average weighted shortest distance.
    
    5. If all pairs of partners are checked, go to 6 otherwise go to 4.
6. End Do.

## RESULTS AND DISCUSSION

Figure 2(a) shows that the algorithm leads to a star network around the founding partners. The partners in Figure 2(a) are coloured according to their individual payoff values, as defined in Eq. (1). The computational partner in the founding network, $P_4$, is the central partner. All new partners create a link to $P_4$, and eight new partners create a link to $P_2$. It is suggested that partners, $P_2$ and $P_4$, are the most convenient partners for hiring experienced researchers and for inviting scientists from outside the network. In addition, Figure 2(b) shows that the individual payoffs of the founding partners are higher than the other partners.

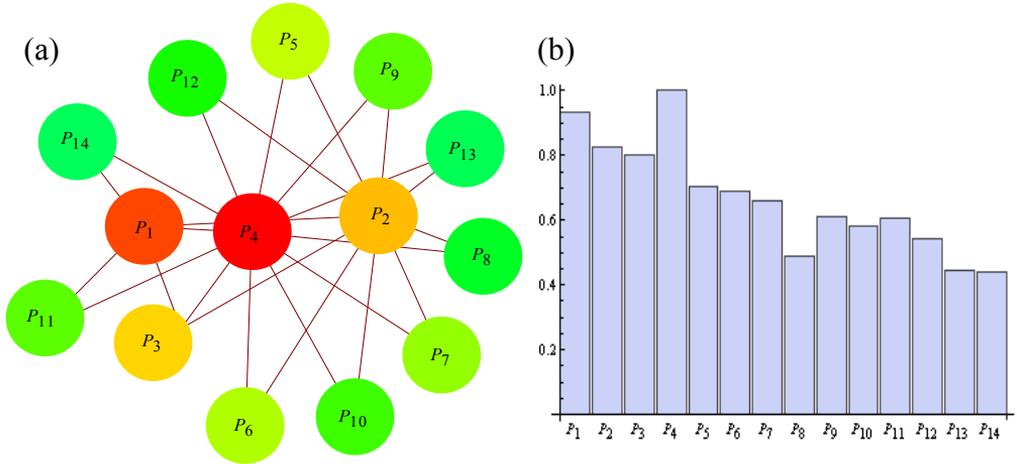

Figure 2. *(a) Graph of the final structure of the network. Partners are coloured according to their individual payoff values in Figure 2(b). (b) Normalised individual payoff values defined in Eq. (1).*

The results are examined using principal component analysis (PCA)[3]. The PCA reveals similar (or different) behaviour among the partners or among the ESRs. Since we have many partners and ESRs, the Euclidian (ordinary) distance has been used for clustering the partners or ESRs. Figure 3 shows that the founding partners and ESRs hired by the founding partners play significant roles in the final network. The method highlights the fact that although partners $P_1$ and $P_3$ have fewer connections compared to other founding members, their payoffs are relatively high. In addition, Figure 3(b) shows that $ESR_2$ and $ESR_4$ play crucial roles because they create the most important connections (associated with longer visits) within the founding network. It can be concluded that the length of visits of ESRs between founding partners are very important for the collaboration of the whole network.

---

[3] PCA is a statistical technique for covariance analysis of multi-dimensional data. The method looks for the largest eigenvalues, and uses them for highlighting the most significant information (or the dominant pattern) hidden in the data. The eigenvectors corresponding to the largest eigenvalues are called the principal components. The first principal component is the eigenvector corresponding to the largest eigenvalue, and the second principal component is the eigenvector corresponding to the second largest eigenvalue.

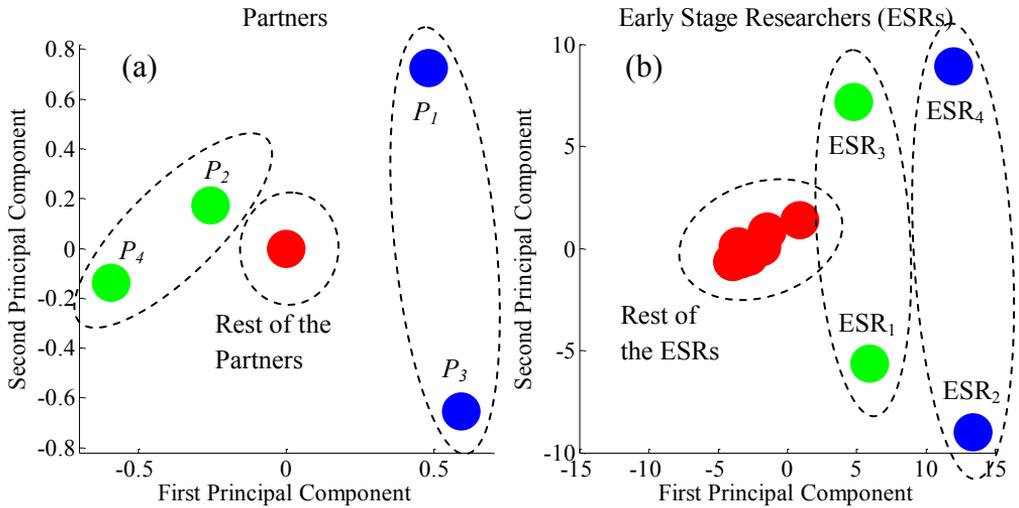

Figure 3. *PCA analysis of final structure of the network. The results are clustered using the Euclidian distance. Different colours are used to highlight the distinct clusters in the final network data. (a) Partners, (b) Early stage researchers (ESRs).*

**ACKNOWLEDGEMENT**

The first author would like to acknowledge the support provided by the European Community's Seventh Framework Programme (ITN-FP7/2007-2013) under the GASMEMS project with a grant agreement no:215504.